\documentclass[a4paper]{jpconf}
\usepackage{graphicx}
\begin{document}
\title{Review of Monte Carlo methods for particle multiplicity evaluation}

\author{N\'estor Armesto}

\address{Department of Physics, CERN, Theory Division, CH-1211 Gen\`eve 23,
Switzerland}

\ead{Nestor.Armesto.Perez@cern.ch}

\begin{abstract}
I present a brief review of the existing models for particle multiplicity
evaluation in heavy ion collisions
which are at our disposal in the form of Monte Carlo simulators.
Models are classified according to the physical mechanisms with which they try
to describe the different stages of a high-energy
collision between heavy nuclei. A comparison
of predictions, as available at the beginning of year 2000,
for multiplicities in central AuAu collisions at the BNL Relativistic Heavy Ion
Collider (RHIC) and PbPb
collisions at the CERN Large Hadron Collider (LHC) is provided.
\end{abstract}

\section{Introduction}

Heavy Ion Physics is a very complex field. Many different physical mechanisms
have to be considered to describe a high-energy collision between heavy
nuclei. The available Monte Carlo tools in the form of simulators,
see the reviews \cite{Bass:1999zq,Armesto:2000xh}, are already
quite sophisticated
but not yet so standardized and developed as the corresponding ones in
hadron-hadron collisions. The reason for this is, of course,
the complexity of the
field.

The central goal of Heavy Ion Physics in the production and
characterization of a new phase of matter, the Quark Gluon Plasma (QGP),
in which deconfinement and chiral
symmetry are restored. In this way
we expect to gain a basic understanding of these still
poorly known, key features of the field theory of strong interactions,
Quantum Chromodynamics (QCD).
But even without the presence of any phase transition,
our understanding of a high-energy
collision involving nuclei is far from being complete.
While we have theoretical control on perturbatively computable quantities,
non-perturbative effects are dealt with
by models. Many of these non-perturbative
effects get enhanced by the nuclear size. The density of produced
particles leads to possible non-linear effects of much less importance in
hadron-hadron collisions. Obviously the possibility of phase transitions makes
the task of describing a heavy ion collisions even more involved. So the
standard procedure for hadron-hadron collisions:
hard collisions (+ radiation) + hadronization + underlying soft event
has to be reconsidered, and many models have appeared through the years. At
variance with hadron-hadron were little attention is paid to the
time evolution of the collision, a space-time picture is
assumed implicitly or explicitly in the models which aim to describe heavy
ion collisions.

The demands to a Monte Carlo event generator are gigantic: they are supposed
to give full information about the event, so they must deal both
with hard probes (jets, heavy flavors,$\dots$),
under theoretical control in perturbative QCD (pQCD), and the underlying soft
event, model-dependent and, at most, inspired in QCD.
Total multiplicities correspond to the latter. Another point which should be
stressed and which makes the task of event generator even more demanding is
the fact that sometimes the underlying
event
is not only the background (for jet quenching, charmonium
suppression,$\dots$)
but the signal itself (strangeness, correlations,
interferometry, flow,$\dots$).
The simplest approach of a superposition of nucleon-nucleon collisions fails,
so some degree of collectivity (or interference) is introduced in most models.
Simple superposition is taken as a baseline. The key question to answer is
whether this collectivity corresponds to a fully thermalized QGP, in which case
the models should provide the initial condition for such state and justify
the way to
thermalization, or to a non-thermalized system. In any of these two cases,
the applicability
of the models has to be seriously examined.

In this review I will consider each simulator as a distinct model. Many
simulators use the same 'physical' model to describe some stage of a heavy ion
collision. But even in this case
both the details of the implementation of the model -- even the
apparently most obvious ones like energy-momentum conservation -- and the
different description of other stages of the collision, vary from model to
model. Besides, model parameters are tuned to 'final stage'
experimental data. As a result the
predictions of the models differ even when they claim to be implementations of
the same physics.

\begin{figure}[htb]
\begin{center}
\includegraphics[width=14cm]{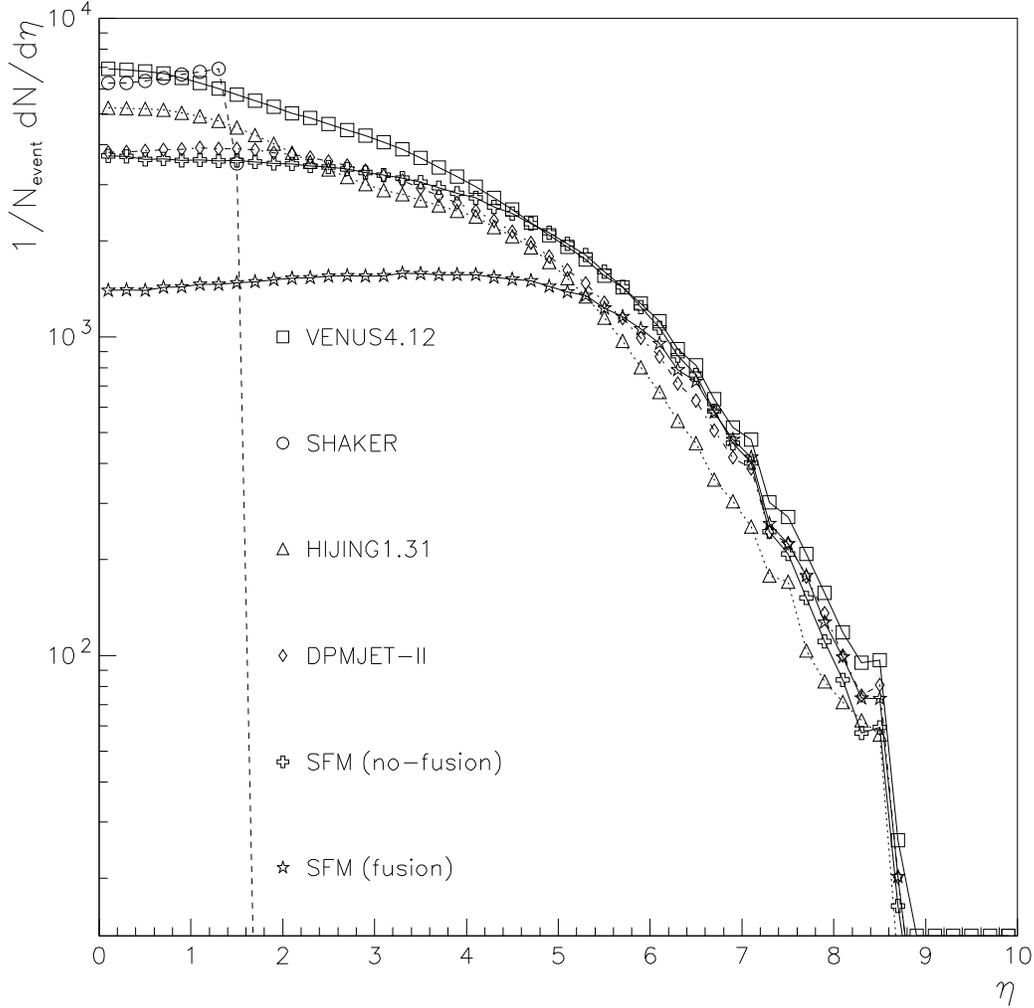}
\end{center}
\caption{\label{fig1}Predictions from different models for the charged
pseudorapidity density in central ($b\le 3$ fm) Pb-Pb collisions at
$\sqrt{s}=6$
TeV per nucleon,
from \protect{\cite{:1995pv}}.}
\end{figure}

Let us have a look to the situation in December 1995 as presented in
\cite{:1995pv}. In Fig. \ref{fig1} differences of a factor $\sim 4$ can be
seen between the model which contains collectivity (lowest line at $\eta=0$)
and all the other models which do not include collective effects. But even
among the latter differences of a factor $\sim 2$ can be observed. Actually it
was this study which led to the strategy of the LHC experiments to make the
simulations in two extreme scenarios, namely 8000 and 2000 charged particles
per unit rapidity at $\eta \simeq 0$ for the background in a central PbPb
event. We will see
that experimental
data coming from RHIC tend to favor those models which the lowest
multiplicities, but uncertainties of a factor $\sim 2$ still exist.

This paper is organized as follows: in the next Section a classification of
models is offered, together with an enumeration and a brief summary of their
main characteristics. Any possible omission has not
been deliberate but it is due to my ignorance, for which I anticipate my
excuses. In Section \ref{pred} predictions from different models for charged
multiplicities in central collisions at RHIC and the
LHC as available in February
2000 \cite{Armesto:2000xh} -- the situation before RHIC started its operation
in June 2000 -- are reviewed. I consider this exercise more
informative than digging into more recent post-dictions which, obviously,
reproduce RHIC data. Finally I present some conclusions.

\begin{figure}[htb]
\begin{center}
\includegraphics[width=14cm]{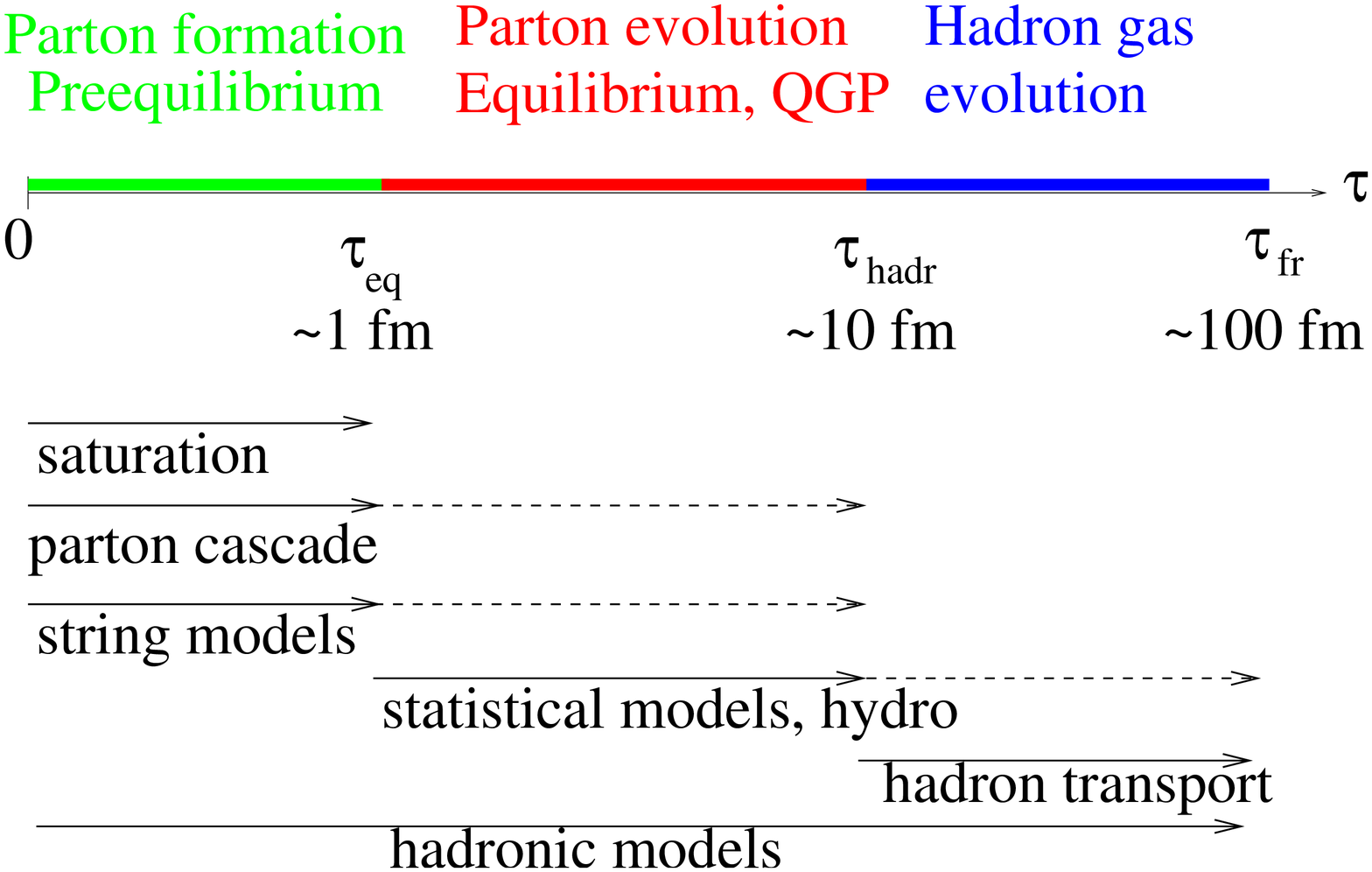}
\end{center}
\caption{\label{fig2}Schematic cartoon of the stages of a heavy ion collision
and of the models which try to describe every stage. Solid lines indicate the
stages for which the models were initially designed. Dashed lines
indicate the stages to which sometimes they are extended. Time scales are
merely indicative of orders of magnitude.}
\end{figure}

\section{Classification of models} \label{clas}

A classification of models for heavy ion collisions is always delicate, due to
the fact that most models are mixed models: They contain different physical
assumptions for each stage of the collision. Here I have tried to use the
stages of a heavy ion collision to offer some guidance (as done in
\cite{Bass:2001gb}), see Fig. \ref{fig2}.
Models will be classified according to what I understand is their main emphasis.
Saturation \cite{Levin:2004ak,marzia}
(no yet available in the form of a Monte Carlo simulator), parton cascade and
string models intend to describe the initial stage, usually called
pre-equilibrium when one assumes that equilibration will occur. They
are further extended to subsequent stages
when no equilibration is assumed. The
equilibrated partonic stage (QGP) is described by hydrodynamics or statistical
models, which can be extended to describe the last stage (hadronic phase).
Hadron transport is only intended to describe this last stage, but hadronic
models exist which try to simulate the whole history of the collision.
For a longer discussion on and description of the different models and their
options, please have a look
to \cite{Armesto:2000xh}.

\subsection{Hadronic models}

Hadronic models consider the description of the nuclear collisions through the
simple superposition of hadron-hadron collisions. The number of such
collisions for a given impact parameter $b$ is determined by nuclear geometry
using the Glauber-Gribov model \cite{Glauber,Gribov:1968jf}.
Both LUCIFER \cite{Kahana:1998wf} and
LEXUS \cite{Jeon:1997bp} belong to this category. The latter is
used as
a baseline: The failure of this approximation means the onset of collective
effects among several nucleon-nucleon collisions.

\subsection{Parton cascade models}

In these models, a relativistic Boltzmann equation is solved for partons. Cross
sections and splittings are computed in pQCD at lowest order. The initial
conditions are provided by either a nuclear wave function like in VNI
\cite{Geiger:1992cm}
or PCM \cite{Shin:2002fg},
or by partons produced in initial parton-parton collisions in
AMPT \cite{Zhang:1999bd} (as
provided by HIJING, see below).
Initial and final state radiation, coherence of the
initial state and the possibility of hadronization via coalescence is included
or under consideration in PCM.

\subsection{String models}

String models describe the collision through the exchange of color
or momentum between partons in the projectile and target. As a consequence of
these exchanges,
these partons become joined by colorless objects, which are called string,
ropes or flux tubes. These models were originally designed for hadron-hadron,
its generalization to hadron-nucleus and nucleus-nucleus done through the
Glauber-Gribov theory \cite{Glauber,Gribov:1968jf}.
Hard collisions (i.e. pQCD computed parton-parton collisions) are included, so
these models contain a soft and a hard component which is crucial for their
application to RHIC and LHC energies. Models of this type are HIJING
\cite{Wang:1991ht},
PSM \cite{Amelin:2001sk}, DPMJET \cite{Ranft:1994fd},
NEXUS \cite{Drescher:1999js} and LUCIAE \cite{Andersson:1996vi}. VENUS
\cite{Werner:1993uh},
Fritiof \cite{Andersson:1986gw} and the SFM \cite{Amelin:1993cs}
are no longer maintained.

Strings are used as initial stage in some transport models (RQMD, UrQMD and
HSD, see below), and as the fragmentation method in most models. Some
collective mechanisms have been introduced, as pomeron interactions in DPMJET
or NEXUS, 
high-color fields or string fusion
in PSM, HIJING, RQMD, LUCIAE
and DPMJET, and
increased stopping through baryon junction migration in HIJING and DPMJET.
Jet quenching has also been introduced in HIJING.

\subsection{Hydrodynamical/statistical models}

In this item I include those models which assume local thermodynamical
equilibrium at a partonic level. Among the models which consider
a hydrodynamical evolution of a QGP, some of them 
emphasize the
phase transition \cite{Bass:1999tu}
to hadronic matter linked to hadronic transport by
UrQMD
while others consider an initial condition given by HIJING or
saturation \cite{Hirano:2002sc}
(introducing also jet quenching inside the plasma). The
statistical models in which hadronization is determined
by phase space, are now
available in Monte Carlo form \cite{becattini}.

\subsection{Hadron transport models}

In these models, a relativistic Boltzmann equation is solved for hadrons in
the final stage of the collision (after hadronization). A huge variety of
hadron species and of cross sections is introduced. These models are AMPT,
RQMD \cite{Sorge:1989dy},
UrQMD \cite{Bass:1998ca} and HSD \cite{Cassing:1997kw}.
Besides, simple models for rescattering of secondaries and spectators
have been introduced in LUCIAE, DPMJET and PSM to explain some
experimental data like multi-strange baryon enhancement.

\subsection{Comments}

Here I will make some comments on the general status of the models:

\begin{itemize}

\item Physically there is no clear border between one stage in a heavy ion
collision and the previous or the subsequent one: the division between the
different stages is indeed artificial. Also this division implies that
causality and some factorization hold. This is assumed but not at all
guaranteed.

\item Energy-momentum conservation is a key demand for all the simulators, but
it is not so easy to fulfill. Usually this means no difficulty for RHIC and LHC
energies, but may be problematic for the most massive nuclei at smaller
energies.

\item 
Models are designed
to describe the full event,
not just one of the stages. Parameters are tuned to experimental data, which
demands some
modeling or assumption about the whole history of the event. So
most models are multi-step ones, containing different pieces corresponding to
the different stages linked one each other.

\item Only DPMJET, HIJING and PSM work for LHC energies (they are included in
the ALICE Generator Pool \cite{alicegp}).

\item Codes are written in Fortran, not in C++. Little standardization exists,
e.g. few of them follow the OSCAR
protocols \cite{oscar}.

\item Pieces of PYTHIA \cite{Sjostrand:1993yb},
ARIADNE \cite{Lonnblad:1992tz} and JETSET
\cite{Andersson:1983ia} are used in
many codes. Charm and beauty production is usually introduced through PYTHIA.

\item Predictions for LHC should be
constrained by RHIC results but are still uncertain in a factor $\sim 2$, due
to existing open questions on the small-$x$ behavior of parton densities, the
degree of collectivity achieved in the collision, the existence or not of
the QGP,$\dots$

\end{itemize}

\section{Predictions for RHIC and LHC (as available in February 2000)}
\label{pred}

A comparison of experimental data with some
available Monte Carlo codes is done in almost every
experimental paper. Many of them can be found in the experimental talks at
this Workshop. For example, \cite{pepe} contains an extensive
comparison of model results for multiplicities at central rapidity
and their centrality evolution at RHIC energies. So instead of looking into
the recent modifications of models to reproduce RHIC data, I will concentrate
in analyzing the situation in February 2000, previous to the first RHIC data
(the first collisions at RHIC happened in June 2000).

Most of the results
presented here come from the review \cite{Armesto:2000xh}, see there for
a description of the models and full references.
In that review results from many models for RHIC and LHC are compiled. But
predictions from the models
have been given for different conditions: centrality cuts done
through impact parameter or percentages of the cross sections, slightly
different energies, AuAu or PbPb collisions both for RHIC and the LHC, $\eta$
or $y\simeq 0$,$\dots$ In order to present them in a compatible manner, I have
corrected each result to the common conditions: 5 \% most central collisions
at $y=0$, for AuAu at 200 GeV per nucleon (RHIC) and PbPb at 5.5 TeV per
nucleon (LHC), using an early version of the Monte Carlo \cite{Amelin:2001sk}.
In \cite{Armesto:2000xh} you can find tables containing the published
predictions of the different models and the correction factors, which were
found to be
as large as 20 \% for RHIC and 17 \% for LHC. This procedure could only be
done properly inside each model, so a systematic
uncertainty of $10\div 15$ \% has to be
understood.

Predictions from seventeen models will be presented in the following
Subsections. Not all of them are available both at RHIC and at the LHC. Let me
briefly indicate the models which ones are contained in the following figures:

\begin{enumerate}

\item DPM \cite{Capella:1999kv}:
a version of the Dual Parton Model including shadowing
corrections due to pomeron interaction diagrams.

\item DPMJET \cite{Ranft:1994fd}.

\item SFM: an early version of \cite{Amelin:2001sk} which is an evolution of
\cite{Amelin:1993cs}.

\item RQMD \cite{Sorge:1989dy}.

\item HIJING \cite{Wang:1991ht}: versions of HIJING with and without jet
quenching give differences for multiplicities
of about a factor 2 for LHC; the value I indicate
in an average of the results with and without quenching.

\item Eskola {\it et al.} \cite{Eskola:1999fc}:
a model containing saturation.

\item HIJING+ZPC+ART, now called AMPT \cite{Zhang:1999bd}.

\item UrQMD \cite{Bass:1998ca}.

\item VNI+UrQMD as contained in \cite{Bass:1999zq}.

\item Hydro+UrQMD \cite{Bass:1999tu}.

\item VNI+HSD as contained in \cite{Bass:1999zq}.

\item VENUS \cite{Werner:1993uh}.

\item NEXUS \cite{Drescher:1999js}.

\item Statistical \cite{Bass:1999zq}: obtained from the results of the
statistical model presented in that reference.

\item WA98 extrapolation \cite{Peitzmann:1999qd}:
extrapolation of low energy
results done by the WA98 Collaboration.

\item WNM: a simple
extrapolation \cite{Armesto:2000xh} to nucleus-nucleus from nucleon-nucleon
done in the Wounded Nucleon Model.

\item Percolation: a simple
estimate \cite{Armesto:2000xh} in the framework of percolation of strings
\cite{Armesto:1996kt}.

\end{enumerate}

I will mainly discuss the results of those models available as Monte Carlo
simulators.

\subsection{RHIC}

\begin{figure}[htb]
\begin{center}
\includegraphics[width=14cm]{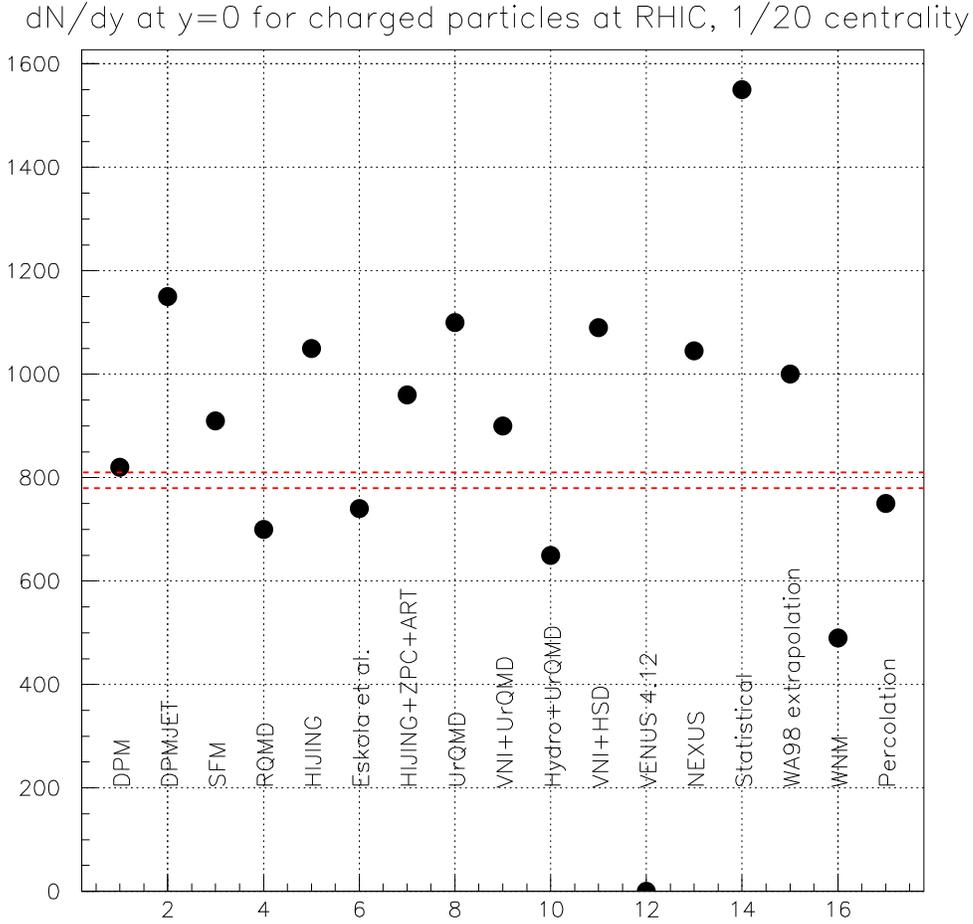}
\end{center}
\caption{\label{fig3}Results from different models for charged particle
multiplicity per unit rapidity at $y=0$ for 5 \% most central AuAu collisions
at 200 GeV per nucleon. Models are explained in the text. The red lines are
only to guide the eye and correspond to experimental data, see the text.}
\end{figure}

In Fig. \ref{fig3} the results from different models for RHIC (charged
particle
multiplicity per unit rapidity at $y=0$ for 5 \% most central AuAu collisions
at 200 GeV per nucleon)
are presented and
compared with experimental data. The red lines define a band where
the central values
for charged multiplicities provided by RHIC experiments lie
(see \cite{pepe} and
references therein). These experimental values have been corrected in the same
way, previously explained, as the results from different models. 

It can be observed that low multiplicities are favored, which indicates a high
degree of coherence or interference. Models without collectivity
tend to give too large values. In order to lower the multiplicities,
now HIJING
has included a large gluon shadowing \cite{Li:2001xa} and DPMJET
considers percolation of strings \cite{Bopp:2004xn}.

\subsection{LHC}

\begin{figure}[htb]
\begin{center}
\includegraphics[width=14cm]{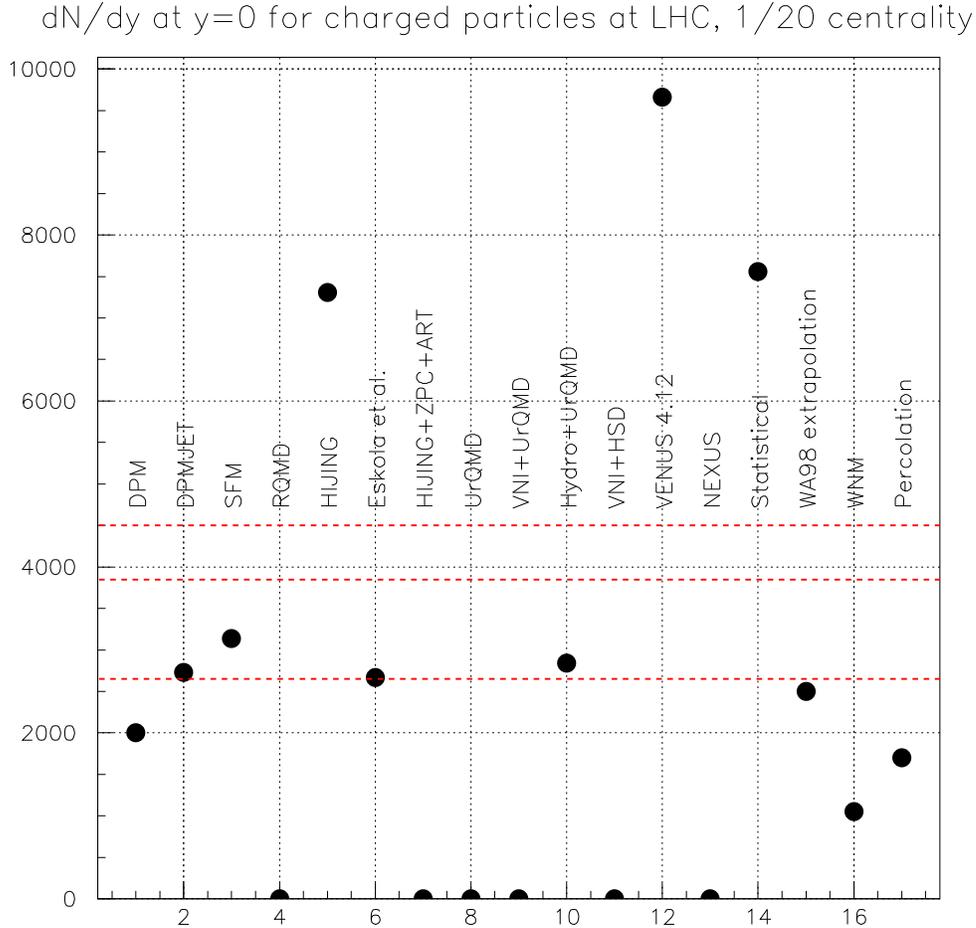}
\end{center}
\caption{\label{fig4}Results from different models for charged particle
multiplicity per unit rapidity at $y=0$ for 5 \% most central PbPb collisions
at 5.5 TeV per nucleon. Models are explained in the text. The
red lines are
only to guide the eye and correspond to averages of models, see the text.}
\end{figure}

In Fig. \ref{fig4} the results from different models for LHC (charged particle
multiplicity per unit rapidity at $y=0$ for 5 \% most central PbPb collisions
at 5.5 TeV per nucleon) are presented. The red lines correspond to averages of
the results of models: The top line corresponds to
all models but WNM and percolation (which are most
naive estimations). The line in the middle corresponds to
all models but WNM and
percolation, and VENUS (which is no longer maintained and gave the largest
prediction already in \cite{:1995pv}, see Fig. \ref{fig1}). And the lower line
corresponds to
all models but WNM, percolation and VENUS,
and HIJING and statistical (in order to produce some kind of average for the
models predicting
the lowest multiplicities, which seem to be favored by RHIC data).

It can be seen that low multiplicities, $2000\div 4000$, are now
favored. But even with RHIC constraints, uncertainties of a factor $\sim 2$
persist, as anticipated. Saturation
\cite{Levin:2004ak,marzia} and percolation of strings
\cite{DiasdeDeus:2000cg,Braun:2001us}
predict multiplicities $\leq
2000$.
On the other hand, predictions from HIJING without quenching, $\sim 5000$, and
with quenching, $\sim 9000$, vary almost a factor 2.

\section{Conclusions} \label{concl}

In this paper I present a brief review of the existing Monte Carlo tools to
simulate heavy ion collisions at ultra-relativistic energies. Some of them:
PSM, DPMJET and HIJING,
are already
available \cite{alicegp} for heavy ion
collisions at the LHC. Let me present some personal comments:

\begin{itemize}

\item With constraints coming from RHIC, low
multiplicities at LHC, $2000\div 4000$, seem to be favored.

\item Collectivity is a well established phenomenon,
but practical realizations
differ both conceptually and quantitatively.

\item  At RHIC and LHC energies, a description of the initial stage in
terms of partons, not hadrons, looks unavoidable.

\item Constraints from RHIC reduce
basic uncertainties for LHC but some of them still remain, e.g.
multiplicity reduction or transverse momentum spectra. This is mainly due to
the fact that RHIC is pre-asymptotic
compared with theory and with the energy available at the LHC.

\item Producing a Monte Carlo simulator demands a large effort which extends
through several years. Some of the existing models are or will be ruled out by
RHIC data, others will
simply extinguish. The interplay between simulator builders
and experimentalists is crucial for the former to keep on creating and
updating their codes. While new models are under production for hadron-hadron
at the LHC, to my knowledge no new model is currently under construction for
the corresponding heavy ion collisions.

\item The actual problems for RHIC are mainly transverse momentum
distributions in AuAu collisions compatible with those in dAu (see
\cite{Bopp:2004xn,Pierog:2003xk}), and elliptic flow (see
\cite{Zabrodin:2001rz,Bopp:2004ip}).

\item Models contain different decoupled
stages. A better theoretical understanding of such decoupling assumption is
needed.

\item Codes are not standardized (compared to hadron-hadron simulators).
The lack of
a reliable, generally accepted
theoretical baseline makes it difficult to establish long
term programs.
Increasing modularity as proposed by the OSCAR protocols \cite{oscar}, with
the possibility to link different initial stages to different models
for
thermalization or intermediate transport and hadronic cascades, would offer new
opportunities for model building.

\item Interplay with the Cosmic Ray Physics community is possible
and highly desirable \cite{karel}.

\item Hard probes like jet quenching,
for which we have reliable
theoretical understanding, must be included in a simulation of the
whole event. Even with a rough knowledge of the underlying event, this would
be a great help.
Also saturation and phase transitions like that from QGP to hadronic matter
or percolation have to be considered as options.

\item The problems of detailed balance
and
Lorentz invariance in parton cascades are still open questions. For
example, the inclusion of $2\to 3$ processes is argued \cite{Xu:2004mz}
to make a huge
difference in thermalization times within a parton cascade.

\end{itemize}

Let me finally comment that looking from outside our field, the
status of Monte Carlo event generators for heavy ion collisions may look still
at the beginning. There are many open fundamental
questions which are theoretically accessible but not yet solved completely.
But it should not be forgotten that not all interesting questions demand such
precision as is intended in
hadron-hadron. Even simple
implementations of the underlying event may provide valuable conclusions. And,
plasma or no plasma,
there is a lot to learn which is equally valuable for hadron-hadron.

\ack
I express my gratitude to the organizers for their invitation to
such a nice meeting. Discussion during the workshop with E. M. Levin, K.
Redlich, K. \u{S}afa\u{r}ik, R. Ugoccioni and many others are gratefully
acknowledged.


\section*{References}

\begin{thebibliography}{99}

\bibitem{Bass:1999zq}
Bass S A {\it et al.} 1999
Last call for RHIC predictions
{\it Nucl. Phys.} A {\bf 661} 205 ({\it Preprint}
nucl-th/9907090)

\bibitem{Armesto:2000xh}
Armesto N and Pajares C 2000
Central rapidity densities of charged particles at RHIC and LHC
{\it Int. J. Mod. Phys.} A {\bf 15} 2019 ({\it Preprint}
hep-ph/0002163)

\bibitem{:1995pv} ALICE Collaboration 1995
ALICE: Technical proposal for a large ion collider experiment at the  CERN
LHC {\it Preprint}
CERN-LHCC-95-71

\bibitem{Bass:2001gb}
Bass S A 2001
Microscopic reaction dynamics at SPS and RHIC
{\it Nucl. Phys.} A {\bf 698} 164
({\it Preprint} nucl-th/0104040)

\bibitem{Levin:2004ak}
Levin E 2004
CGC, QCD saturation and RHIC data (Kharzeev-Levin-McLerran-Nardi point of
view)
({\it Preprint} hep-ph/0408039, these Proceedings)

\bibitem{marzia} Nardi M 2004 (these Proceedings)

\bibitem{Glauber} Glauber R J 1959 Lectures on Theoretical Physics Vol. 1
Brittin W E and Duham L G eds. (New York, Interscience)

\bibitem{Gribov:1968jf}
Gribov V N 1969
Glauber Corrections And The Interaction Between High-Energy Hadrons And
Nuclei
{\it Sov. Phys. JETP} {\bf 29} 483
[{\it Zh. Eksp. Teor. Fiz.}  {\bf 56} 892]

\bibitem{Kahana:1998wf}
Kahana D E and Kahana S H 1998
Two phase simulation of ultrarelativistic nuclear collisions
{\it Phys. Rev.} C {\bf 58} 3574
({\it Preprint} nucl-th/9804017)

\bibitem{Jeon:1997bp}
Jeon S and Kapusta J I 1997
Linear extrapolation of ultrarelativistic nucleon nucleon scattering to
nucleus nucleus collisions
{\it Phys. Rev.} C {\bf 56} 468
({\it Preprint} nucl-th/9703033)

\bibitem{Geiger:1992cm}
Geiger K 1992
Particle production in high-energy nuclear collisions: A Parton cascade
cluster hadronization model
{\it Phys. Rev.} D {\bf 47} 133

\bibitem{Shin:2002fg}
Shin G R and Muller B 2002
A relativistic parton cascade with radiation
{\it J. Phys.} G {\bf 28} 2643
({\it Preprint} nucl-th/0207041)

\bibitem{Zhang:1999bd}
Zhang B, Ko C M, Li B A and Lin Z w 1999
A multi-phase transport model for nuclear collisions at RHIC
{\it Phys. Rev.} C {\bf 61} 067901
({\it Preprint} nucl-th/9907017)

\bibitem{Wang:1991ht}
Wang X N and Gyulassy M 1991
HIJING: A Monte Carlo model for multiple jet production in p p, p A and A A
collisions
{\it Phys. Rev.} D {\bf 44} 3501

\bibitem{Amelin:2001sk}
Amelin N S, Armesto N, Pajares C and Sousa D 2001
Monte Carlo model for nuclear collisions from SPS to LHC energies
{\it Eur. Phys. J.} C {\bf 22} 149
({\it Preprint} hep-ph/0103060)

\bibitem{Ranft:1994fd}
Ranft J 1994
The Dual parton model at cosmic ray energies
{\it Phys. Rev.} D {\bf 51} 64

\bibitem{Drescher:1999js}
Drescher H J, Hladik M, Ostapchenko S and Werner K 1999
A new approach to nuclear collisions at RHIC energies
{\it Nucl. Phys.} A {\bf 661} 604
({\it Preprint} hep-ph/9906428)

\bibitem{Andersson:1996vi}
Andersson B and Tai A 1996
The Firecracker Model, a possible collective effect in high-energy heavy
ion
collisions
{\it Z. Phys.} C {\bf 71} 155

\bibitem{Werner:1993uh}
Werner K 1993
Strings, pomerons, and the venus model of hadronic interactions at
ultrarelativistic energies
{\it Phys. Rept.}  {\bf 232} 87

\bibitem{Andersson:1986gw}
Andersson B, Gustafson G and Nilsson-Almqvist B 1987
A Model For Low P(T) Hadronic Reactions, With Generalizations To Hadron -
Nucleus And Nucleus-Nucleus Collisions
{\it Nucl. Phys.} B {\bf 281} 289

\bibitem{Amelin:1993cs}
Amelin N S, Braun M A and Pajares C 1993
Multiple production in the Monte Carlo string fusion model
{\it Phys. Lett.} B {\bf 306} 312

\bibitem{Bass:1999tu}
Bass S A, Dumitru A, Bleicher M, Bravina L, Zabrodin E, Stoecker H and
Greiner W 1999
Hadronic freeze-out following a first order hadronization phase  transition
in ultrarelativistic heavy-ion collisions
{\it Phys. Rev.} C {\bf 60} 021902
({\it Preprint} nucl-th/9902062)

\bibitem{Hirano:2002sc}
Hirano T and Nara Y 2002
Energy loss in high energy heavy ion collisions from the hydro+jet
model
{\it Phys. Rev.} C {\bf 66} 041901
({\it Preprint} hep-ph/0208029)

\bibitem{becattini} Becattini F 2004 (these Proceedings)

\bibitem{Sorge:1989dy}
Sorge H, Stocker H and Greiner W 1989
Poincare Invariant Hamiltonian Dynamics: Modeling Multi - Hadronic
Interactions In A Phase Space Approach
{\it Annals Phys.}  {\bf 192} 266

\bibitem{Bass:1998ca}
Bass S A {\it et al.} 1998
Microscopic models for ultrarelativistic heavy ion collisions
{\it Prog. Part. Nucl. Phys.}  {\bf 41} 225
({\it Preprint} nucl-th/9803035)

\bibitem{Cassing:1997kw}
Cassing W and Bratkovskaya E L 1997
Production and absorption of c anti-c pairs in nuclear collisions at  SPS
energies
{\it Nucl. Phys.} A {\bf 623} 570
({\it Preprint} hep-ph/9705257)

\bibitem{alicegp} ALICE Collaboration 2002
ALICE Physics Performance Report, Chapter IV:
Monte
Carlo generators and simulations Carminati F, Foka Y,
Giubellino P,
Paic G, Revol J-P, \u{S}afa\u{r}ik K and Wiedemann U A eds.
{\it Preprint} ALICE Internal Note 2002-033

\bibitem{oscar} http://nt3.phys.columbia.edu/OSCAR/

\bibitem{Sjostrand:1993yb}
Sjostrand T 1993
High-energy physics event generation with PYTHIA 5.7 and JETSET 7.4
{\it Comput. Phys. Commun.} {\bf 82} 74

\bibitem{Lonnblad:1992tz}
Lonnblad L 1992
ARIADNE version 4: A Program for simulation of QCD cascades implementing
the
color dipole model
{\it Comput. Phys. Commun.}  {\bf 71} 15

\bibitem{Andersson:1983ia}
Andersson B, Gustafson G, Ingelman G and Sjostrand T 1983
Parton Fragmentation And String Dynamics
{\it Phys. Rept.} {\bf 97} 31

\bibitem{pepe} Milov A 2004 (these Proceedings)

\bibitem{Capella:1999kv}
Capella A, Kaidalov A and Tran Thanh Van J 1999
Gribov theory of nuclear interactions and particle densities at future
heavy-ion colliders
{\it Heavy Ion Phys.}  {\bf 9} 169
({\it Preprint} hep-ph/9903244)

\bibitem{Eskola:1999fc}
Eskola K J, Kajantie K, Ruuskanen P V and Tuominen K 1999
Scaling of transverse energies and multiplicities with atomic number  and
energy in ultrarelativistic nuclear collisions
{\it Nucl. Phys.} B {\bf 570} 379
({\it Preprint} hep-ph/9909456)

\bibitem{Peitzmann:1999qd}
Peitzmann T {\it et al.}  [WA98 Collaboration] 1999
Recent results from the WA98 experiment
{\it Nucl. Phys.} A {\bf 661} 191

\bibitem{Armesto:1996kt}
Armesto N, Braun M A, Ferreiro E G and Pajares C 1996
Percolation approach to quark-gluon plasma and J/psi suppression
{\it Phys. Rev. Lett.}  {\bf 77} 3736
({\it Preprint} hep-ph/9607239)

\bibitem{Li:2001xa}
Li S y and Wang X N 2001
Gluon shadowing and hadron production at RHIC
{\it Phys. Lett.} B {\bf 527} 85
({\it Preprint} nucl-th/0110075)

\bibitem{Bopp:2004xn}
Bopp F W, Ranft J, Engel R and Roesler S 2004
RHIC data and the multichain Monte Carlo DPMJET-II
{\it Preprint} hep-ph/0403084

\bibitem{DiasdeDeus:2000cg}
Dias de Deus J and Ugoccioni R 2000
Particle densities in heavy ion collisions at high energy and the dual
string model
{\it Phys. Lett.} B {\bf 491} 253
({\it Preprint} hep-ph/0008086)

\bibitem{Braun:2001us}
Braun M A, Del Moral F and Pajares C 2001
Percolation of strings and the first RHIC data on multiplicity and
transverse
momentum distributions
{\it Phys. Rev.} C {\bf 65} 024907
({\it Preprint} hep-ph/0105263)

\bibitem{Pierog:2003xk}
Pierog T, Drescher H J, Liu F M, Ostapchenko S and Werner K 2003
High p(T) suppression without jet quenching in Au + Au collisions in
NEXUS
{\it Preprint} hep-ph/0307048

\bibitem{Zabrodin:2001rz}
Zabrodin E E, Fuchs C, Bravina L V and Faessler A 2001
Elliptic flow at collider energies and cascade string models: The role  of
hard processes and multi-pomeron exchanges
{\it Phys. Lett.} B {\bf 508} 184
({\it Preprint} nucl-th/0104054)

\bibitem{Bopp:2004ip}
Bopp F W, Ranft J, Engel R and Roesler S 2004
Learning from RHIC data with DPMJET-III
{\it Acta Phys. Polon.} B {\bf 35} 303

\bibitem{karel} \u{S}afa\u{r}ik K 2004 (these Proceedings)

\bibitem{Xu:2004mz}
Xu Z and Greiner C 2004
Thermalization of gluons in ultrarelativistic heavy ion collisions by
including three-body interactions in a parton cascade
{\it Preprint} hep-ph/0406278


\end{thebibliography}
\end{document}